**Plasmon-polaritons and diffraction on the layer of asymmetric hyperbolic metamaterial**

**M. V. Davidovich**[1,a]
[1]*Department of Physics, Saratov State University, Astrakhanskaya 83, 410012, Saratov, Russia*
[a] Electronic mail: davidovichmv@info.shu.ru

We consider the plasmon-polaritons along a layer of hyperbolic metamaterial propagating in the plane of the anisotropy axis with an arbitrary its orientation. As a layer material, we use periodic plane-layered artificial medium – hyperbolic metamaterial of thin metal and dielectric layers and produce its homogenization. The conditions for the existence of fast, slow, leakage, gliding flowing, forward and backward plasmon-polaritons are found. The Fresnel formulas for the diffraction of a plane wave of arbitrary polarization on such a structure are obtained. The dispersion of plasmon-polaritons and plane wave diffraction are calculated. It is proposed to use a strong magnetic field to control dispersion and scattering.

**1. Introduction**

In recent years there is increased interest in metamaterials or artificial media (AM) with hyperbolic dispersion law [1–22], called hyperbolic metamaterials (HMM), including the HMM in the frequency area where one of the components of the effective dielectric permittivity (DP) tensor is closed to zero, or the so-called ENZ (epsilon-near-zero) AM [23]. The HMM is usually a single-axis electromagnetic AM or photonic crystal (PC), where homogenization gives different signs of the real part of the two main diagonal components of the tensor of effective DP. It is usually made of thin conductive metal, semiconductor or graphene layers periodically embedded in the dielectric background Fig. 1. In such a uniaxial electromagnetic crystal, the axis is directed perpendicular to the layers, and two transverse to the axis components of the DP tensor may have the property $\varepsilon'_\perp < 0$, $\varepsilon_\perp = \varepsilon'_\perp - i\varepsilon''_\perp$, whereas for the longitudinal component $\varepsilon'_{||} > 0$, $\varepsilon_{||} = \varepsilon'_{||} - i\varepsilon''_{||}$ (HMM of the second type). The HMM of the first type is usually made of conductive nanocylinders (nanowires) periodically embedded in the dielectric background [1–6]. For them, the condition $\varepsilon'_{||} = \varepsilon'_{zz} < 0$ is possible where the axis of the HMM is directed along the axis of the cylinders. We will consider the HMM of the second type as a plane-layered periodic AM or PC, but show that under some conditions it behaves as a HMM of the first type. Let the medium consist of thin metal layers of nanoscale thickness $t_m$, periodically embedded with a period $t_p = t_m + t_d$ in a non-dissipative dielectric with DP $\varepsilon_d > 1$. Here $t_d$ is the distance between the layers of metal or the thickness of the dielectric layer Fig. 1. The DP of metal we take in the Drude-Lorentz form $\varepsilon_m = \varepsilon_L - \omega_p^2/(\omega^2 - i\omega\omega_c)$, or $\varepsilon_m = \varepsilon' - i\varepsilon''$.

In reality, the DP of thin layers depends on their thickness and is determined by quantum effects. We will use approximate parameters for a massive or bulk silver sample with $\omega_p = 1.5 \cdot 10^{16}$ Hz, $\omega_c = 4.1 \cdot 10^{13}$ Hz, $\varepsilon_L = 9$. Let the $z$ axis is directed perpendicular to the layers. The value $\varepsilon' = \varepsilon_L - \omega_p^2/(\omega^2 + \omega_c^2)$ will be negative for frequencies $\omega < \sqrt{\omega_p^2/\varepsilon_L - \omega_c^2}$ or for wavelengths $\lambda$ of about 350 nm, and the value $\varepsilon'' = \omega_p^2 \omega_c/(\omega^3 + \omega\omega_c^2)$ is small. Let consider the conditions of weak dissipation. Obviously, it is $\omega_c \ll \omega < \sqrt{\omega_p^2/(\varepsilon_L+1) - \omega_c^2} \approx \omega_p/\sqrt{\varepsilon_L+1}$. This is a condition for the existence of slow plasmon- polaritones (PP) at the metal-vacuum boundary. In this case $\varepsilon' < -1$, $\varepsilon''/|\varepsilon'| \ll 1$. The waves of arbitrary directions are investigated in the HMM when the propagation direction does not coincide with the axis [21]. In the case of waveguide formation as a layer, such a HMM is called asymmetric [9]. Recently, waves in waveguides with a dielectric core and a shell of HMM have been studied [24,25]. In these works the dissipation was not taken into account. In this paper, we investigate a waveguide as a layer of asymmetric HMM with an arbitrary orientation of the anisotropy axis Fig.1 and when considering dissipation.

## 2. Method of research

For the considered AM in the approximation of the absence of spatial dispersion (SD) the homogenization is given by simple formulas [13]

$$\varepsilon_\perp = \varepsilon_{xx} = \varepsilon_{yy} = (t_m \varepsilon_m + t_d \varepsilon_d)/t_p, \quad \varepsilon_\parallel = \varepsilon_{zz} = \left((t_m/t_p)\varepsilon_m^{-1} + (t_d/t_p)\varepsilon_d^{-1}\right)^{-1}.$$

The formulas taking into account the SD can be found in [13] and in a number of other works. Then we think $\varepsilon_d$ of the order of 2–5. Let find the condition, when $\varepsilon'_\perp < 0$. Denoting the filling factor of the metal $K = t_m/t_d$, we obtain

$$\omega < \sqrt{\frac{\omega_p^2}{(\varepsilon_L + (1/K - 1)\varepsilon_d)} - \omega_c^2}.$$

If we neglect the dissipation and take $K = 0.5$, then will be $\omega < \omega_p/\sqrt{12}$, i.e. the wave lengths are more than 400 nm. To perform homogenization in the optical range, it is sufficient to use structures with period $t_p < 40$ nm, i.e. with layer thicknesses of the order of 20 nm or less. For simplicity, we put $K = 0.5$ and obtain the parameters of effective DP for weak dissipation. Consider a few cases. Suppose first that $\varepsilon' = -\varepsilon_d$. In this case $\varepsilon_{xx} = -i\varepsilon''/2$, $\varepsilon_{zz} = 2\varepsilon_d(1 - i\varepsilon_d/\varepsilon'')$, i.e. the transverse component is small and imaginary, and the longitudinal component is highly dissipative. Let now $\varepsilon' << -\varepsilon_d$. Then

$$\varepsilon_{xx} = (\varepsilon' + \varepsilon_d - i\varepsilon'')/2,$$

$$\varepsilon'_{zz} = \frac{2\varepsilon_d}{(|\varepsilon'| - \varepsilon_d)}\left(|\varepsilon'| + \varepsilon''^2/(|\varepsilon'| - \varepsilon_d)\right), \quad \varepsilon''_{zz} = \frac{2\varepsilon_d \varepsilon''}{(|\varepsilon'| - \varepsilon_d)}\left(|\varepsilon'|/(|\varepsilon'| - \varepsilon_d) - 1\right).$$

If $-\varepsilon_d < \varepsilon' < 0$ and $\varepsilon_d - |\varepsilon'_m| >> \varepsilon''$, то $\varepsilon'_{xx} = (\varepsilon_d - |\varepsilon'|)/2 > 0$, that we have

$$\varepsilon'_{zz} = \frac{2\varepsilon_d}{(\varepsilon_d - |\varepsilon'|)}\left(-|\varepsilon'| + \varepsilon''^2/(\varepsilon_d - |\varepsilon'|)\right), \quad \varepsilon''_{zz} = \frac{2\varepsilon_d \varepsilon''}{(\varepsilon_d - |\varepsilon'|)}\left(1 + |\varepsilon'|/(\varepsilon_d - |\varepsilon'|)\right).$$

In this case, provided $|\varepsilon'| > \varepsilon''^2/(\varepsilon_d - |\varepsilon'|)$, we obtain $\varepsilon'_{xx} > 0$ and $\varepsilon'_{zz} < 0$, i.e. the AM becomes the HMM of the first kind. Let finally $\varepsilon' \approx 0$. In this region at finite dissipation we have $\varepsilon_{zz} = 2\varepsilon''(\varepsilon''/\varepsilon_d - i)$, $\varepsilon_{xx} = (\varepsilon_d - i\varepsilon'')/2$, i.e. the longitudinal component is small and strongly dissipative. This is the so-called ENZ region [23]. We will further be interested in the case $\varepsilon'_{zz} = -\varepsilon'_{xx}$. This equation is easily solved. If $\varepsilon' << -\varepsilon_d$, then the condition $\varepsilon'_{zz} = -\varepsilon'_{xx}$ is met to within the small term $\varepsilon''^2/(2\varepsilon'^2)$ if $\varepsilon_d = (3 - \sqrt{8})\varepsilon'|$ or $\varepsilon' = -\varepsilon_d(3 + 2\sqrt{2})$. In this case we have $\varepsilon'_{zz} = -\varepsilon'_{xx} + \varepsilon''^2/(2\varepsilon'^2)$, $\varepsilon''_{zz} = \varepsilon''\left(1/2 - \sqrt{3 - \sqrt{8}}\right) = 0.086\varepsilon''$. If $-\varepsilon_d < \varepsilon' < 0$, then the condition $\varepsilon'_{zz} = -\varepsilon'_{xx}$ leads to the solution $\varepsilon_d = (3 + \sqrt{8})\varepsilon'|$. In this case one can again find $\varepsilon'_{zz} = -|\varepsilon'_{xx}| + \varepsilon''^2/(2\varepsilon'^2)$, and $\varepsilon''_{zz} = \varepsilon''\left(\sqrt{3 + \sqrt{8}} + (3 + \sqrt{8})/2\right) = 5.32\varepsilon''$, i.e. the dissipation here is higher. Obviously, when the frequency is shifted, the condition $\varepsilon'_{zz} = -\varepsilon'_{xx}$ can be fulfilled accurately, and the dissipation will somewhat change.

In an infinite medium of HMM, we consider an electromagnetic wave of the form $\mathbf{E}(x, z, t) = \mathbf{E}_0 \exp(i(\omega t - k_x x - k_z z))$, $\mathbf{H}(x, z, t) = \mathbf{H}_0 \exp(i(\omega t - k_x x - k_x z))$. This medium is described by a homogenized effective permittivity tensor

$$\hat{\varepsilon} = \begin{bmatrix} \varepsilon_{xx} & 0 & 0 \\ 0 & \varepsilon_{yy} & 0 \\ 0 & 0 & \varepsilon_{zz} \end{bmatrix}, \tag{1}$$

in which for HMM of periodic plane-layered structures with conductive films, the normal to which is oriented along the $z$ axis, we have $\varepsilon_{xx} = \varepsilon_{yy} = \varepsilon'_\perp - i\varepsilon''_\perp$, and the value $\varepsilon'_\perp$ can be negative. Consider the matrix of rotation of the structure around the $y$ axis by an angle $\alpha$:

$$\hat{T}(\alpha) = \begin{bmatrix} \cos(\alpha) & 0 & -\sin(\alpha) \\ 0 & 1 & 0 \\ \sin(\alpha) & 0 & \cos(\alpha) \end{bmatrix}. \tag{2}$$

Acting on vector **E**, it gives: $E'_x = E_x \cos(\alpha) - E_z \sin(\alpha)$, $E'_y = E_y$, $E'_z = E_x \sin(\alpha) + E_z \cos(\alpha)$, i.e. there is a counterclockwise rotation. The matrix (1) will take the form $\tilde{\varepsilon} = \hat{T}^{-1}(\alpha)\hat{\varepsilon}\hat{T}(\alpha) = \hat{T}(-\alpha)\hat{\varepsilon}\hat{T}(\alpha)$ or

$$\tilde{\varepsilon} = \begin{bmatrix} \cos^2(\alpha)\varepsilon_{xx} + \sin^2(\alpha)\varepsilon_{zz} & 0 & \sin(\alpha)\cos(\alpha)(\varepsilon_{zz} - \varepsilon_{xx}) \\ 0 & \varepsilon_{xx} & 0 \\ \sin(\alpha)\cos(\alpha)(\varepsilon_{zz} - \varepsilon_{xx}) & 0 & \cos^2(\alpha)\varepsilon_{zz} + \sin^2(\alpha)\varepsilon_{xx} \end{bmatrix}. \tag{3}$$

We write the homogeneous Maxwell equations in such AM as $\nabla \times \mathbf{H} = i\omega\varepsilon_0 \tilde{\varepsilon}\mathbf{E}$, $\nabla \times \mathbf{E} = -i\omega\mu_0 \mathbf{H}$. Painting them by components, we have

$$\partial_y H_z - \partial_z H_y = -\partial_z H_y = i\omega\varepsilon_0(\tilde{\varepsilon}_{xx} E_x + \tilde{\varepsilon}_{xz} E_z),$$

$$\partial_z H_x - \partial_x H_z = i\omega\varepsilon_0 \tilde{\varepsilon}_{xx} E_y,$$

$$\partial_x H_y - \partial_y H_x = \partial_x H_y = i\omega\varepsilon_0(\tilde{\varepsilon}_{xz} E_x + \tilde{\varepsilon}_{zz} E_z),$$

$$\partial_y E_z - \partial_z E_y = -\partial_z E_y = -i\omega\mu_0 H_x,$$

$$\partial_z E_x - \partial_x E_z = -i\omega\mu_0 H_y,$$

$$\partial_x E_y - \partial_y E_x = \partial_x E_y = -i\omega\mu_0 H_z.$$

In these equations, we took into account that the fields are independent of $y$. These equations are divided into two systems of equations: $E_y \neq 0$, $H_y = 0$ and $H_y \neq 0$, $E_y = 0$. The first has the form:

$$Z_0(k_x H_z - k_z H_x) = k_0 \tilde{\varepsilon}_{xx} E_y,$$

$$k_z E_y = -Z_0 k_0 H_x,$$

$$k_x E_y = Z_0 k_0 H_z,$$

$$H_y = E_x = E_z = 0.$$

The second, in which $E_y = H_x = H_z = 0$, is written as

$$Z_0 k_z H_y = k_0(\tilde{\varepsilon}_{xx} E_x + \tilde{\varepsilon}_{xz} E_z),$$

$$-Z_0 k_x H_y = k_0(\tilde{\varepsilon}_{xz} E_x + \tilde{\varepsilon}_{zz} E_z),$$

$$k_z E_x - k_x E_z = Z_0 k_0 H_y.$$

Here $Z_0 = c\mu_0 = \sqrt{\mu_0/\varepsilon_0}$. The first type of equations gives the H-wave with respect to $z$-axis, and the second one is E-wave. Consider first the latter. Equality to zero of its determinant gives Fresnel's dispersion equation (DE) $\tilde{\varepsilon}_{xx} k_x^2 + \tilde{\varepsilon}_{zz} k_z^2 + 2\tilde{\varepsilon}_{xz} k_x k_z = k_0^2(\tilde{\varepsilon}_{xx}\tilde{\varepsilon}_{zz} - \tilde{\varepsilon}_{xz}^2)$. From two equations

$$\tilde{\varepsilon}_{xx} E_x + \tilde{\varepsilon}_{xz} E_z = Z_0 H_y k_z / k_0,$$

$$\tilde{\varepsilon}_{xz} E_x + \tilde{\varepsilon}_{zz} E_z = -Z_0 H_y k_x / k_0$$

the following relations may be written:

$$E_x = \frac{Z_0 H_y}{k_0} \frac{\tilde{\varepsilon}_{zz} k_z + \tilde{\varepsilon}_{xz} k_x}{\tilde{\varepsilon}_{xx}\tilde{\varepsilon}_{zz} - \tilde{\varepsilon}_{xz}^2}, \quad E_z = -\frac{Z_0 H_y}{k_0} \frac{\tilde{\varepsilon}_{xz} k_z + \tilde{\varepsilon}_{xx} k_x}{\tilde{\varepsilon}_{xx}\tilde{\varepsilon}_{zz} - \tilde{\varepsilon}_{xz}^2}. \tag{4}$$

Substituting them in the third, we have the DE, which we write in the form

$$\frac{k_x^2}{\tilde{\varepsilon}_{zz}} + \frac{k_z^2}{\tilde{\varepsilon}_{xx}} + \frac{2\tilde{\varepsilon}_{xz}}{\tilde{\varepsilon}_{xx}\tilde{\varepsilon}_{zz}} k_x k_z = k_0^2 \frac{\Delta}{\tilde{\varepsilon}_{xx}\tilde{\varepsilon}_{zz}}. \tag{5}$$

Here it's marked $\Delta = \tilde{\varepsilon}_{xx}\tilde{\varepsilon}_{zz} - \tilde{\varepsilon}_{xz}^2$. From this Fresnel equation for the extraordinary wave the two values are defined

$$k_x^\pm = -k_z \tilde{\varepsilon}_{xz}/\tilde{\varepsilon}_{xx} \pm \sqrt{(k_0^2 \tilde{\varepsilon}_{xx} - k_z^2)\Delta}/\tilde{\varepsilon}_{xx}. \tag{6}$$

These two values correspond to opposite waves along $\pm x$. The second equation in (4) allows us to find the impedance:

$$Z^\pm = -E_z/H_y = \frac{Z_0}{k_0} \frac{\tilde{\varepsilon}_{xz} k_z + \tilde{\varepsilon}_{xx} k_x^\pm}{\Delta} = \pm Z_0 \sqrt{(\tilde{\varepsilon}_{xx} - k_z^2/k_0^2)/\Delta}. \tag{7}$$

It depends on the direction: $Z^\pm = \pm Z_0 \rho$, $\rho = \sqrt{(\tilde{\varepsilon}_{xx} - k_z^2/k_0^2)/\Delta}$. Equating it to the impedance of an E-wave in a vacuum $Z_0 \rho_0 = Z_0 \sqrt{1 - k_z^2/k_0^2}$ propagating along the $x$-axis, we obtain the dispersion equation (DE) for the E-plasmon-polariton (EPP) along the surface in $z$ direction

$$\rho = \sqrt{1 - k_z^2/k_0^2}. \tag{8}$$

It defines two waves along each direction: $k_z = \pm k_0 \sqrt{(\Delta - \tilde{\varepsilon}_{xx})/(\Delta - 1)}$. In the case of symmetry $\tilde{\varepsilon}_{xz} = 0$ we have the solution $k_z = \pm k_0 \sqrt{\tilde{\varepsilon}_{xx}(\tilde{\varepsilon}_{zz} - 1)/(\tilde{\varepsilon}_{xx}\tilde{\varepsilon}_{zz} - 1)}$. This PP very slow, if $\tilde{\varepsilon}_{xx}\tilde{\varepsilon}_{zz} \approx 1$ (Fig. 2). In the case $\tilde{\varepsilon}_{xx} = \tilde{\varepsilon}_{zz}$ this is Zenneck DE [26–30]. In the layered structure this equality is impossible. Such waveguide must be a homogeneous layer, either dielectric or metallic. In the latter case, the maximum deceleration will be at $\varepsilon' = -1$, i.e. at the frequency of the plasmonic resonance.

In the case of a wave in the plate, the solution is

$$H_y = \exp(-ik_z z)\left[A^+ \exp(-ik_x^+ x) + A^- \exp(-ik_x^- x)\right], \tag{9}$$

the components of the electric field are determined by the formulas (4). We'll need a component

$$E_z = \frac{\exp(-ik_z z)}{-c\varepsilon_0 k_0}\left[A^+ \rho \exp(-ik_x^+ x) - A^- \rho \exp(-ik_x^- x)\right]. \tag{10}$$

In vacuum, we also need the solutions of the wave equation in the form:

$$\begin{aligned} H_y &= B\exp(-ik_z z)\exp(-ik_{0x}(x-d)), \\ E_z &= -BZ_0(k_{0x}/k_0)\exp(-ik_z z)\exp(-ik_{0x}(x-d)), \\ H_y &= C\exp(-ik_z z)\exp(ik_{0x}x), \\ E_z &= CZ_0(k_{0x}/k_0)\exp(-ik_z z)\exp(ik_{0x}x). \end{aligned} \tag{11}$$

They are written for areas $x > d$ and $x < 0$ respectively. Here we have $k_{0x}^2 + k_z^2 = k_0^2$. Moreover, the direction of energy motion of the fast wave ($\mathrm{Re}(k_z^2) < k_0^2$) is taken from the plate into vacuum (leakage). For a symmetric structure $k_x^+ = -k_x^-$. The fields on both sides have identical dependencies, and can have either an exponential decay in the vacuum side (surface gliding wave), or an exponential growth (antisurface or leakage wave). Therefore, it is enough to enter one constant in (11) with matching the fields on one surface [28]. The gliding means the movement of energy from the vacuum on both sides and absorption it in the plate. At weak energy inflow the surface wave can be weakly dissipative. Leakage means the emission of stored energy from the plate into the vacuum. Strong leakage even with weak dissipation is accompanied by large radiation losses. Signs in (11) are chosen according to the conditions of emission, i.e. outflow of energy. Leakage can be replaced by gliding with increasing frequency. Gliding from one side and leakage from the other side (or in the other direction) for the structure under consideration are not possible. For the symmetric case, the proof is simple. Its DE is obtained by equating the input impedance on the one hand to the impedance of the wave in vacuum [26–30], i.e. imposing the condition $R = 0$. By transforming the impedance of a wave into vacuum and equating it to the same impedance, we obtain $\tan(k_x d) = 0$. This is not a DE, but the matching condition in which the reflection coefficient is zero. For the transparent layer, this is the condition of the matching energy output for the transmission line at half-wave thickness of the dielectric layer. Transformation means unidirectional energy transfer. The DE is produced, if we change the sign of one of the impedances. For the asymmetric case the different leakage/gliding conditions lead to a change in the sign of the impedance and $k_{0x}$ in one of the equations (11). Such system of equations has no solution. Determining $k_{0x} = \pm\sqrt{k_0^2 - k_z^2}$, it should be taken into account that the sign should be chosen so that the slow wave in the dissipative structure of the HMM from the vacuum was gliding with energy flowing ($k'_{0x} < 0$), i.e. the energy from the vacuum should flow into the plate [26–30]. In this sense, taking $k_{0x} = k'_{0x} - ik''_{0x}$, we should require the implementation $k'_{0x} < 0$ and $k''_{0x} > 0$ for gliding wave and $k'_{0x} > 0$, $k''_{0x} < 0$ for leakage one. One can see that the gliding wave is surface (decreasing from the surface towards vacuum) and the leakage wave is antisurface (exponentially increasing). Let take $k_z = k'_z - ik''_z$. At weak dissipation in a slow wave $k'^2_z > k_0^2$ we have

$$k_{0x} = \mp i\sqrt{k_z^2 - k_0^2} = \mp i\sqrt{k'^2_z - k''^2_z - k_0^2 - 2ik'_z k''_z} \approx$$
$$\approx \mp i\sqrt{k'^2_z - k_0^2 - 2ik'_z k''_z} \approx \mp i\sqrt{k'^2_z - k_0^2}\left(1 - \frac{ik'_z k''_z}{k'^2_z - k_0^2}\right),$$

therefore $k'_{0x} = -\sqrt{k'^2_z - k_0^2}$, $k''_{0x} = -k'_z k''_z / \sqrt{k'^2_z - k_0^2}$, i.e. the minus sign is taken, and the wave is really gliding. From above, the wave falls from the vacuum at an angle of leakage $\theta = \arctan(k'_{0x}/k'_z) = -\arctan\left(\sqrt{1 - k_0^2/k'^2_z}\right)$ (Fig. 1), and from below – at an angle $-\theta$. In case the wave is fast. For it $k_{0x} = \pm\sqrt{k_0^2 - k'^2_z - k''^2_z + 2ik'_z k''_z} \approx \pm\sqrt{k_0^2 - k'^2_z}\left(1 + ik'_z k''_z/(k_0^2 - k'^2_z)\right)$. In this case, it should be taken the "plus" sign, $k'_{0x} = \sqrt{k_0^2 - k'^2_z}$, $k''_{0x} = k'_z k''_z/\sqrt{k_0^2 - k'^2_z}$, and the wave is leaking, Fig. 1 at the leakage angle $\theta = \arctan(k'_x/k'_z) = \arctan\left(\sqrt{k_0^2/k'^2_z - 1}\right)$. It should be noted that the direction of the wave along $z$ we determine as the direction of motion of energy, i.e. for a positive we take such, when $k''_z > 0$. Therefor it is the dependence $\exp(-k''_z z)$, i.e. attenuation in the direction of motion of energy. If at the same time $k'_z > 0$, such a wave is forward. If, however $k'_z < 0$, wave is backward. In it the phase and the energy move oppositely. We see that in the backward wave with weak dissipation, the energy flow is replaced by the outflow and vice versa. However, with strong dissipation, the root extraction may not lead to such an effect, i.e. all modes are possible. Matching tangential components, we get

$$B = A^+ \exp(-ik_x^+ d) + A^- \exp(-ik_x^- d),$$
$$C = A^+ + A^-,$$
$$B\rho_0 = \rho[A^+ \exp(-ik_x^+ d) - A^- \exp(-ik_x^- d)], \quad (12)$$
$$-C\rho_0/\rho = A^+ - A^-.$$

Here $A^{\pm} = C(1 \mp \rho_0/\rho)$. We divide the third equation by the first and get DE in the form

$$\rho_0 = \rho \frac{(\rho - \rho_0)\exp(-ik_x^+ d) - (\rho + \rho_0)\exp(-ik_x^- d)}{(\rho - \rho_0)\exp(-ik_x^+ d) + (\rho + \rho_0)\exp(-ik_x^- d)}. \quad (13)$$

In the case of symmetry it takes the form

$$\rho_0 = -\rho \frac{\rho_0 + i\rho \tan(k_x d)}{\rho + i\rho_0 \tan(k_x d)},$$

or $2i\rho_0\rho = (\rho^2 + \rho_0^2)\tan(k_x d)$. There are the solutions $\rho_0 = \rho[i\tan(k_x d/2)]^{\pm 1}$. The upper sign corresponds to the electric wall in the center, and the lower one corresponds to the magnetic wall. In generally

$$2\rho\rho_0\alpha = \rho^2 + \rho_0^2, \quad \alpha = \frac{\exp(-ik_x^+ d) + \exp(-ik_x^- d)}{\exp(-ik_x^+ d) - \exp(-ik_x^- d)}.$$

There are two DEs: $\rho_0 = \rho(\alpha \pm \sqrt{\alpha^2 - 1})$. Let denote $\beta_{\pm} = \alpha \pm \sqrt{\alpha^2 - 1}$. Then for the square of the deceleration we find $n^2 = (\tilde{\varepsilon}_{xx}\tilde{\varepsilon}_{zz} - \tilde{\varepsilon}_{xz}^2 - \tilde{\varepsilon}_{xx}\beta_{\pm}^2)/(\tilde{\varepsilon}_{xx}\tilde{\varepsilon}_{zz} - \tilde{\varepsilon}_{zx}^2 - \beta_{\pm}^2)$. Large deceleration is possible if the denominator is small or the numerator is large.

Let consider the DE for H-PP. Substituting the components of the magnetic field in the first equation, we have the DE $k_x^2 + k_z^2 = k_0^2 \tilde{\varepsilon}_{xx}$. This is the Fresnel equation for ordinary wave. For this wave the impedance along the $x$-axis has the form $Z = E_y/H_z = Z_0 k_0/\sqrt{k_0^2 \tilde{\varepsilon}_{xx} - k_z^2}$. It cannot be matched with the corresponding impedance in vacuum, so H-PP along the boundary plane of infinite HMM sample does not exist. However, it occurs if there is a finite layer of HMM. In this case, as well as for a single metal layer [26–35], we have two solutions:

$$\sqrt{k_0^2 \varepsilon_{xx} - k_z^2} = \sqrt{k_0^2 - k_z^2} \left[i \tan\left(d\sqrt{k_0^2 \tilde{\varepsilon}_{xx} - k_z^2}/2\right)\right]^{\pm 1}. \quad (14)$$

The solution with the sign "plus" corresponds to the electric wall, and with the sign "minus" – to the magnetic wall in the center of the layer. These equations are the same as for the metal layer with replacement $\varepsilon_{xx} \to \varepsilon_m$ [26,29]. Since slow PPs $k_z'^2 > k_0^2$ are possible, it is convenient the DE (14) to convert:

$$\sqrt{k_z^2 - k_0^2 \tilde{\varepsilon}_{xx}} = \sqrt{k_z^2 - k_0^2} \left[\tanh\left(d\sqrt{k_z^2 - k_0^2 \tilde{\varepsilon}_{xx}}/2\right)\right]^{\mp 1}. \quad (15)$$

Denote the hyperbolic tangent as $T$ in the square equation (15) and assume that the plasmon is sufficiently slow. Then we have $k_z^2 = k_0^2(1 - \tilde{\varepsilon}_{xx} T^{\mp 2})/(1 + T^{\mp 2})$. If dissipation can be neglected, then $k_z^2 = k_0^2(1 + |\tilde{\varepsilon}_{xx}'| T^{\mp 2})/(1 - T^{\mp 2})$. In this case, for a slow PP the value $T$ must be less than one and close to it. We see that it is possible the solution $k_z^- = k_0\sqrt{(1 + |\tilde{\varepsilon}_{xx}'| T^2)/(1 - T^2)}$ with slow PP, and the plasmon with $k_z^+$ cannot be very slow and weakly dissipative.

### 3. Investigation of special cases

Consider some special cases. They occur at different angles of the optical axis with the $z$ axis or between the normal to the surface and the planes of the layers. Case $\alpha = 0$. For it $\tilde{\varepsilon}_{xz} = 0$ and $k_x^{\pm} = \pm k_x = \pm\sqrt{\varepsilon_{zz}(k_0^2 - k_z^2/\varepsilon_{xx})}$. You can write the DE for a layer as

$$k_z^2/k_0^2 = \varepsilon_{xx} \frac{\varepsilon_{zz} + \tan^{\pm 2}(k_x d/2)}{\tan^{\pm 2}(k_x d/2) + \varepsilon_{xx}\varepsilon_{zz}}.$$

Let $\varepsilon' = 0$. Then $\varepsilon_{xx} = (\varepsilon_d - i\varepsilon'')/2$, the longitudinal component $\varepsilon_{zz} = -2i\varepsilon''\varepsilon_d/(\varepsilon_d - i\varepsilon'') \approx -2i\varepsilon''/\varepsilon_d$ is small and highly dissipative. The ordinary wave in this HMM behaves as in the dielectric. For an unusual wave under these conditions, an approximate solution for PP along the surface of a semi-infinite sample follows from (8)

$$k_z = \pm k_0 \sqrt{\varepsilon_d}\left(1 - i\varepsilon''(\varepsilon_d + 1/\varepsilon_d)/2\right). \quad (16)$$

This is a slow PP along the surface of an infinitely thick sample. For him from the Fresnel equation follows $k_x = (1 - i)k_0\sqrt{\varepsilon''/\varepsilon_d}/2$, i.e. it fades deep into the sample. Also we have

$$k_{0x} = \sqrt{k_0^2 - k_0^2 \varepsilon_d\left(1 - \varepsilon''^2(\varepsilon_d + 1/\varepsilon_d)^2/4 - i\varepsilon''(\varepsilon_d + 1/\varepsilon_d)/2\right)} \approx -ik_0\sqrt{\varepsilon_d - 1},$$

i.e. the PP decays in the vacuum. Consider now the frequency area $\varepsilon' = -\varepsilon_d$. For it $\varepsilon_{xx} = -i\varepsilon''/2$, $\varepsilon_{zz} = 2\varepsilon_d(1 - i\varepsilon_d/\varepsilon'')$. From DE (8) it is follows

$$k_z = \pm k_0 \frac{\varepsilon_d}{\sqrt{1 + \varepsilon_d^2}}\left(1 + i\varepsilon'' \frac{\varepsilon_d + (\varepsilon_d - 1)(1 + \varepsilon_d^2)}{4\varepsilon_d^2(1 + \varepsilon_d^2)}\right).$$

This is backward fast weakly dissipative PP. The DE (13) is convenient to analyze in the case of small layer thickness. In this case, replacing the tangent with its argument we have

$$\varepsilon_{xx}\varepsilon_{zz}\left(1 - k_z^2/k_0^2\right) = \left(k_z^2/k_0^2 - \varepsilon_{xx}\right)\left[\varepsilon_{zz}\left(1 - k_z^2/k_0^2/\varepsilon_{xx}\right)(k_0 d/2)^2\right]^{\pm 1}. \quad (17)$$

Let consider the square bracket small. Also consider that the same way $k_0 d \ll 1$. In the case of the upper sign the PP fast. Its dispersion is determined from the quadratic equation. Replacing in the right side the $k_z^2/k_0^2$ by the unit, approximately in the range of small frequencies, we get $k_z^2/k_0^2 = 1 + (1 - \varepsilon_{xx})^2 (k_0 d/(2\varepsilon_{xx}))^2$. Here it is assumed that the value $|\varepsilon_{xx}|$ is not small. In the case of the lower sign (magnetic wall) the PP is slower, and at low frequencies will be $k_z \approx k_0\sqrt{\varepsilon_{xx}}$. Denote $T = i\tan(k_x d/2) = \tanh\left(d\sqrt{\varepsilon_{zz}(k_z^2 - \varepsilon_{xx}k_0)}\right) = 1 - \delta$. If $|\varepsilon_{zz}| \neq 0$, in the region of high slowing down the hyperbolic tangent is close to unity, i.e. the $\delta$ is a small quantity. Therefore

$$k_z^2/k_0^2 \approx \varepsilon_{xx}\frac{\varepsilon_{zz} - 1}{\varepsilon_{xx}\varepsilon_{zz} - 1} \pm 2\delta\frac{\varepsilon_{xx}\varepsilon_{zz}(\varepsilon_{xx} - 1)}{(\varepsilon_{xx}\varepsilon_{zz} - 1)^2}.$$

The condition of slowness here is the same as for the half-plane: it is a plasmon resonance $\varepsilon_{xx}\varepsilon_{zz} \to 1$. Near it $|\delta| < |\varepsilon_{xx}\varepsilon_{zz} - 1| \ll 1$, as the hyperbolic tangent tends to unity exponentially fast.

*Corner* $\alpha = \pm\pi/2$. This case corresponds to the lower Fig. 1, i.e. the structure turns into a plane-layered waveguide. Its rigorous analysis without homogenization is possible [13]. The PP is possible in the region $\varepsilon'_{xx} < 0$ and it similar to the PP along the metal layer. This case is interesting in terms of comparing the strict dispersion solution with the one obtained on the basis of homogenization and will be discussed below.

*Corner* $\alpha = \pm\pi/4$. In this case, the matrix (3) takes the form

$$\tilde{\varepsilon} = \frac{1}{2}\begin{bmatrix} (\varepsilon_{xx} + \varepsilon_{zz}) & 0 & \pm(\varepsilon_{zz} - \varepsilon_{xx}) \\ 0 & \varepsilon_{xx} & 0 \\ \pm(\varepsilon_{zz} - \varepsilon_{xx}) & 0 & (\varepsilon_{zz} + \varepsilon_{xx}) \end{bmatrix}. \quad (18)$$

It is simplified if $\varepsilon'_{zz} = -\varepsilon'_{xx} > 0$, since then $\tilde{\varepsilon}_{xx} = \tilde{\varepsilon}_{zz} \approx -0.3i\varepsilon''$, $\tilde{\varepsilon}_{xz} \approx \pm(\varepsilon'_{zz} + i\varepsilon''/4)$. The other case $\varepsilon'_{zz} = -\varepsilon'_{xx} < 0$ leads to the values $\tilde{\varepsilon}_{xx} = \tilde{\varepsilon}_{zz} = -2.91i\varepsilon''$, $\tilde{\varepsilon}_{xz} \approx \pm(-\varepsilon'_{xx} - i2.41\varepsilon'')$. In this case, the DP tensor (18) is more dissipative. In the absence of dissipation the DE (5) takes the form $k_x k_z = \mp k_0^2 \tilde{\varepsilon}_{xz}/2$ and determines the asymptotes of the hyperbolic law of dispersion. In this approximation, the components of the wave vector are not limited. The DE for PP along the border of an infinitely thick sample of HMM will now take the form of

$$k_z^2 / k_0^2 = \frac{\varepsilon_{xx} + \varepsilon_{zz} - 2\varepsilon_{xx}\varepsilon_{zz}}{2(1 - \varepsilon_{xx}\varepsilon_{zz})}. \tag{19}$$

In the above frequency band this PP is fast:

$$k_z^2 / k_0^2 \approx \frac{\varepsilon_d^2(3 + 2\sqrt{2})}{1 + \varepsilon_d^2(3 + 2\sqrt{2})}.$$

From Fresnel's equation (6) we have $k_x^+ = -k_0^2 \tilde{\varepsilon}_{xx} / 2k_z$, $k_x^- = -2k_z(1 - k_0^2 \tilde{\varepsilon}_{xx}/(4k_z^2))$. Let consider the DE (13) and PP along the layer. In the considered area $\beta_\pm^2 = -[1 \pm \cos(k_z d)]^2 / \sin^2(k_z d)$, therefore PP cannot be slow. Fast waves at a nonzero angle are associated with the PP running along the metal layers at an angle to the axis. Because $\varepsilon_{xx}\varepsilon_{zz} = \varepsilon_m\varepsilon_d$, the PP at a given angle can be slow if $\varepsilon_m \approx 1/\varepsilon_d$, i.e. above the frequency $\omega_p / \sqrt{\varepsilon_L}$. It is already high-frequency optical polaritone. The solution of the question of the forward or backward wave is obtained by calculating $k_z' = k_z - ik_z''$ and determining the sign of the relation $k_z'/k_z''$ [26–35]. This value $k_z'$ determines the movement phase and the damping $k_z''$ determines the movement of energy in the direction of attenuation. Therefore, $k_z'/k_z'' > 0$ means the forward wave, and $k_z'/k_z'' < 0$ corresponds to backward one. Another way to determine is to calculate the component of the Poynting vector for an unusual wave

$$S_z = \text{Re}(E_x H_y^*)/2 = \frac{Z_0 |H_y|^2}{2k_0} \text{Re}\left(\frac{\tilde{\varepsilon}_{zz} k_z + \tilde{\varepsilon}_{xz} k_x}{\tilde{\varepsilon}_{xx}\tilde{\varepsilon}_{zz} - \tilde{\varepsilon}_{xz}^2}\right) \tag{20}$$

and a similar component for an ordinary wave

$$S_z = \text{Re}(E_y H_x^*)/2 = Z_0^{-1} |E_y|^2 k_z'/(2k_0). \tag{21}$$

Note that here $|H_y(x,z)|^2 = |H_y(x,0)|^2 \exp(-2k_z'' z)$ and $|E_y(x,z)|^2 = |E_y(x,0)|^2 \exp(-2k_z'' z)$. Immediately we see that for the ordinary wave and H-PP the direction of power flow gives the value $k_z'$. In vacuum, the ordinary wave corresponds to the *s*-polarization and the same component (21) of the Poynting vector. For *p*-polarization in vacuum $S_z = \text{Re}(E_x H_y^*)/2 = Z_0 |H_y|^2 k_z'/(2k_0)$, therefore, determining the direction of energy movement for E-PP requires the solution of DE and the calculation of the integral from $S_z$ over infinite cross section, which is a more complicated procedure. In addition, the formula (20) is written for an infinite sample, and then it is possible to substitute any of the two values $k_x^\pm$ of (6). For the final sample, two waves in the transverse direction should be taken into account and their determined amplitudes (see below), which further complicates the task. Therefore, we will use the method based on the definition the sign of the value $k'_z/k''_z$. His discomfort is manifested only in the transition region from fast to slow waves, where $k'_z \approx \pm k_0$ and $k''_z \approx 0$. The equivalence of both approaches is proved for simple waveguide structures of the metal strip type. To solve the problem of gliding (inflow) or leakage (outflow) waves, the component of the Poynting vector $S_x$ in vacuum should be calculated at the value $k_z$ obtained from the DE. This is more simple procedure. For the upper half-space with E-PP we have $S_x = -\mathrm{Re}(E_z H_y^*)/2 = Z_0 |H_y|^2 k'_x/(2k_0)$, i.e. when $k'_x > 0$ the wave is leakage and when $k'_x < 0$ it is gliding. The relations (11) are written so that there is a simultaneous gliding or leakage in both half-spaces. The gliding/leakage conditions change with frequency. We can consider the case when the layer is surrounded by half-spaces of different materials and gliding/leakage conditions are opposite in both half-spaces. In this case, the existing symmetry disappears, and it is possible for the energy from one half-space to flow into the layer and outflow from the layer to the other half-space. In the case of a layer in a vacuum, this is possible only if an external constant magnetic field is applied, which will be discussed further. For such a solution it is necessary to change the sign either in the first two or in the last two equations of the system (11). This results to four-value solution $k_z$ of DE.

*Let* $\alpha = \pi/2$. Again we consider dissipation as small value, and $K = 0.5$. Note that with a different fill factor $K$, the ratios become more complicated. If homogenization is used taking into account spatial dispersion, then all components of tensors become complex functions of $k_x$ and $k_z$, therefore, the analytical study becomes extremely difficult. Now the components are swapped, and the tensor takes the form

$$\tilde{\varepsilon} = \begin{bmatrix} \varepsilon_{zz} & 0 & 0 \\ 0 & \varepsilon_{xx} & 0 \\ 0 & 0 & \varepsilon_{xx} \end{bmatrix}. \tag{22}$$

There are $\varepsilon_{zz} = (\varepsilon_m + \varepsilon_d)/2$ and $\varepsilon_{xx} = 2(\varepsilon_m \varepsilon_d)/(\varepsilon_m + \varepsilon_d)$. This case differs from the case of $\alpha = 0$ simply replacing the components. It is possible to construct a strict model for it without homogenization. Namely, we introduce the dielectric layer transfer matrix

$$\hat{a}_d = \begin{bmatrix} \cos\left(t_d \sqrt{k_0^2 \varepsilon_d - k_z^2}\right) & i\rho_d \sin\left(t_d \sqrt{k_0^2 \varepsilon_d - k_z^2}\right) \\ i\sin\left(t_d \sqrt{k_0^2 \varepsilon_d - k_z^2}\right)/\rho_d & \cos\left(t_d \sqrt{k_0^2 \varepsilon_d - k_z^2}\right) \end{bmatrix},$$

and the similar matrix $\hat{a}_m$ of the metal layer by substitution $d \to m$, $\varepsilon_d \to \varepsilon_m$. Here $\rho_d = \sqrt{(\varepsilon_d - k_z^2/k_0^2)/(1 - k_z^2/k_0^2)}/\varepsilon_d$ is normalized to vacuum impedance the E-wave impedance in the dielectric. We will consider symmetric structures that begin with a metal layer and end with it, or begin with a dielectric layer and end with it. In the first case, the number of metal layers per unit more, and in the second case – the number of dielectric layer per unit more. In the first case, the structure matrix has the form $\hat{a} = (\hat{a}_m \hat{a}_d)^n \hat{a}_m$, and in the second case $\hat{a} = \hat{a}_d (\hat{a}_m \hat{a}_d)^n$, where $n$ is the number of complete periods. If the Fresnel problem is solved for such a structure, the relations take place

$$1 + R = \hat{a}_{11} T + \hat{a}_{12} T, \qquad (23)$$

$$(1 - R)/\rho_0 = (\hat{a}_{21} T + \hat{a}_{22} T)/\rho_0, \qquad (24)$$

which lead to the solution $T = 2/(\hat{a}_{11} + \hat{a}_{22} + \hat{a}_{12} + \hat{a}_{21})$, $R = (Z_{in} - 1)/(Z_{in} + 1)$, $Z_{in} = (\hat{a}_{11} + \hat{a}_{12})/(\hat{a}_{21} + \hat{a}_{22})$. Obviously, the result depends on the symmetry of the structure, i.e. whether it ends in a metallic or dielectric layer. The value $Z_{in}$ is the input impedance of the structure, when the output wave is emitted into vacuum. To obtain DE of PP it should be noted that the direction of energy transfer in (23) and (24) is taken from left to right (for fig. 1 from bottom to top, i.e. the wave falls along the $x$-axis). In a symmetrical structure, either gliding or leakage waves are possible. They are obtained by matching impedances, i.e. by no reflection: $R = 0$. Therefore, in the ratio (24), either on the left or on the right, the sign $\rho_0$ should be changed. This leads to DE $\hat{a}_{11} + \hat{a}_{12} + \hat{a}_{21} + \hat{a}_{22} = 0$. In the case of one dielectric layer it has the form

$$1 + i(\rho_d + 1/\rho_d)\tan\left(t_d \sqrt{k_0^2 \varepsilon_d - k_z^2}\right)/2 = 0.$$

It can be divided into two, using the half-argument tangent formula and solving the quadratic equation. They have a known form [28]. The same equation is obtained using transformation the impedance $-\rho_0 = -\sqrt{1 - k_z^2/k_0^2}$ by the layer and equating it to the impedance $\rho_0$. In the case of a large number of periods, the structure acquires the properties of photonic crystal. For him, the Floquet-Bloch equation has the form

$$\cos(k_x(k_0, k_x)t) = X = (a_{11} + a_{22})/2. \qquad (25)$$

Here the elements of the matrix of one period are taken, i.e. either $a = a_d a_m$, or $a = a_m a_d$. As a result, we find $k_x^\pm = \pm \arccos(X) + 2n\pi$. Also we will use $n = 0$. Note that the Fresnel equation gives $k_x^\pm = \pm \sqrt{k_0^2 \varepsilon_{xx} - k_z^2 \varepsilon_{xx}/\varepsilon_{zz}}$. Taking the matrix $a = \hat{a}_d \hat{a}_m$ of period and considering the smallness of the period compared to the wavelength, we have $X = 1 - t_d t_m \sqrt{k_0^2 \varepsilon_d - k_z^2} \sqrt{k_0^2 \varepsilon_m - k_z^2} (\rho_d / \rho_m + \rho_m / \rho_d)/2$.

Assuming $K = 0.5$ we get $X = 1 - u$ where $u = (tk_0)^2 \sqrt{\varepsilon_d - k_z^2/k_0^2} \sqrt{\varepsilon_m - k_z^2/k_0^2} (\rho_d / \rho_m + \rho_m / \rho_d)/8$.

Because of the smallness $(tk_0)^2$ and $u$, finally we get $k_x^\pm = \pm k_x$ where $k_x = k_0 \sqrt{u/2}$. Let consider a very slow PP. For him $\rho_d = 1/\varepsilon_d$, $\rho_m = 1/\varepsilon_m$, $u = (tk_0)^2 k_z^2 / k_0^2 (\rho_d / |\rho_m| + |\rho_m|/\rho_d)/8$, $k_x^\pm = \pm t k_0 k_z \sqrt{\varepsilon_d / |\varepsilon_m| + |\varepsilon_m|/\varepsilon_d}/4$. Here we took into account that the DP of the metal is approximately negative. The formulas are valid if $t|k_z| \ll 1$. This homogenization more accurately describes the structure. Now you can build a solution and find the dispersion equation. Inside the structure, we also look for it in the form (9), but taking into account the ratio $k_x^\pm = \pm k_x$. This allows you to record

$$H_y = A^\pm \begin{pmatrix} \cos(k_x x) \\ \sin(k_x x) \end{pmatrix} \exp(-ik_z z),$$

express the remaining components through $H_y$ and construct an even $H_y(x)$ solution (with an electric wall in the plane $x = 0$) and an odd $H_y(x)$ solution (with a magnetic wall). It has the form (14) and can be represented as

$$\sqrt{k_0^2 - k_z^2} = \frac{k_x}{\varepsilon_{xx}} [i \tan(k_x d / 2)]^{\pm 1}, \quad (26)$$

where $k_x = k_0 \sqrt{u/2}$. If the Fresnel equation is used instead of the Floquet-Bloch equation, that $k_x = \sqrt{k_0^2 \varepsilon_{xx} - k_z^2 \varepsilon_{xx}/\varepsilon_{zz}}$, and then

$$k_z = \pm k_0 \sqrt{\frac{1 - T^{\pm 2}/\varepsilon_{xx}}{1 - T^{\pm 2}/(\varepsilon_{xx} \varepsilon_{zz})}} = \pm k_0 \sqrt{\frac{1 + 2T^{\pm 2}/(|\varepsilon'| - \tilde{\varepsilon}_d)}{1 + T^{\pm 2}/(\varepsilon_d(|\varepsilon'| + i\varepsilon''))}}.$$

This PP may be slow in the region $|\varepsilon'| \approx \varepsilon_d$, but it is dissipative. A similar equation for the case $\alpha = 0$ has the form

$$\sqrt{k_0^2 - k_z^2} = \frac{k_x}{\varepsilon_{zz}} [i \tan(k_x d / 2)]^{\pm 1}. \quad (27)$$

Here $k_x = \sqrt{k_0^2 \varepsilon_{zz} - k_z^2 \varepsilon_{zz}/\varepsilon_{xx}}$. In these two cases, there is a symmetry in $x$. For (27)

$$k_z = \pm k_0 \sqrt{\frac{1 - T^{\pm 2}/\varepsilon_{zz}}{1 - T^{\pm 2}/(\varepsilon_{xx} \varepsilon_{zz})}} = \pm k_0 \sqrt{\frac{1 - T^{\pm 2}(\tilde{\varepsilon}_d - |\varepsilon'|)/(2\varepsilon_m \varepsilon_d)}{1 - T^{\pm 2}/(\varepsilon_d \varepsilon_m)}}.$$

This PP will be slow in an area $\varepsilon' = 0$ with the value in this area $k_z = \pm k_0 \sqrt{\varepsilon_d / 2}$.

## 4. HMM layer in an external magnetic field

The external magnetic field $\mathbf{H}_0$ causes to the tensor value of DP of metal. By changing the magnitude and direction of the magnetic field, you can control the dispersion and waves in the layer of HMM. The tensor of the effective DP (1) changes accordingly. At an arbitrary angle $\alpha$ and direction of the magnetic field, the relations become very complex. Consider the case of the direction of the magnetic field along the *x*-axis $\mathbf{H}_0 = \mathbf{x}_0 H_0$ and $\alpha = 0$. In this case for the metal sample

$$\hat{\varepsilon}_m = \begin{bmatrix} \varepsilon_{xx}^m & 0 & 0 \\ 0 & \varepsilon_{zz}^m & -ib \\ 0 & ib & \varepsilon_{zz}^m \end{bmatrix},$$

where $\varepsilon_{xx}^m = 1 - \omega_p^2/(\omega^2 - i\omega\omega_c)$, $\varepsilon_{zz}^m = 1 - \omega_p^2/(\omega^2 - \omega_M^2 - i\omega\omega_c)$, $b = \omega_M \omega_p^2/(\omega^3 - \omega\omega_M^2 - i\omega^2\omega_c)$. Now for the DP tensor of the layered HMM (1) we have $\varepsilon_{xx} = (\varepsilon_{xx}^m + \varepsilon_d)/2$, $\varepsilon_{yz} = -\varepsilon_{zy} = -ib/2$, $\varepsilon_{zz} = 2\varepsilon_{zz}^m \varepsilon_d/(\varepsilon_{zz}^m + \varepsilon_d)$, and it ceases to be diagonal. Such an AM is not formally subject to the definition of HMM, but as in the latter, large values of the components of the vector $\mathbf{k}$ are possible in it [22]. The magnetic field leads to the appearance of the gyration vector and additional spatial dispersion. Painting the coordinates of the Maxwell equation, we obtain the Fresnel equation and the connection of fields through impedances. The Fresnel equation is of the form $\det[\hat{k}^2 - k_0^2 \hat{\varepsilon}] = 0$ where the matrices $\hat{k}$ and $\hat{k}^2$ are containing the components $k_l$ and $k_l k_m$ correspondently [22]. They define the operators "rotor" and "rotor from rotor", i.e. act on an arbitrary plane wave $\mathbf{E}(\omega, \mathbf{k})$ as $\nabla \times \mathbf{E} = \hat{k}\mathbf{E}$ and $\nabla \times \nabla \times \mathbf{E} = \hat{k}^2 \mathbf{E}$. The matrix is $\hat{k}$ singular, and the matrix $\hat{k}^2$ is reversible. The differences now is that the impedances will be different. This leads to the fact that the structure is possible be gliding from one side and leakage from the other side. Changing the direction of the magnetic field changes the gliding/leakage conditions. In this sense, such a waveguide structure has non-reciprocity. By exciting, for example, such layer with a waveguide, you can create an antenna of a leakage wave that radiates in a narrow sector at an angle to one side. By changing the magnetic field, you can change the half-plane into which the radiation takes place. Because $\omega_M = e\mu_0 H_0/m_e$, then for work in optical range you need highly strong magnetic field. If the external magnetic field is directed along the *z*-axis, the DP tensor is modified by replacement $x \leftrightarrow z$. For such a layer, the waves of different directions are different.

Decelerating structures with delayed plasmons are promising for the creation of THz-band traveling wave tubes [36,37]. A deceleration coefficient of the order of 3-4 and a working magnetic field induction of the order of 1 T are required. For this purpose it is convenient to use the considered structures with $\alpha = \pi/2$ and the longitudinal magnetic field for focusing of electron beam. Plasmon deceleration and losses depend on the direction of the magnetic field, so the calculation of dispersion taking into account the magnetic field is very important. In the case of an arbitrarily directed **H** field and an arbitrary orientation $\alpha$, the DP tensor has all the components, and the dispersion equations become very complicated.

### 5. The Fresnel's formulas

The HMM structure is equivalent to a set of plane-parallel waveguides or lattices rotated at an angle and is able to effectively control the diffraction of a plane wave, especially if optically pumped semiconductor layers or graphene sheets are used. Tensor conductivity of graphene from THz to UV ranges was obtained in a number of studies [38–41]. In the first approximation in the Kubo-Greenwood model it can be considered scalar [38]. At low frequencies it is inductive, but taking into account interband transitions, it can have a capacitive region and even becomes negative with external pumping [41]. Possible homogenization, taking into account the tensor character of the conductivity $\sigma_{xx} \neq \sigma_{yy}$, $\sigma_{xy} = \sigma_{yx}$, and scalar conductivity [9,10,42]. However, it should be taken into account that the tensor conductivity of graphene does not divide into E-waves and H-waves, so it is necessary to cross-link all four tangent to the boundaries field components or use the 4x4 transmission matrix [43]. This complicates the DE and Fresnel's formulas. Metal tapes with a small thickness $t_m$ can be considered as surface current density $\mathbf{J} = i\omega\varepsilon_0(\varepsilon_m - 1)t_m\mathbf{E} = \sigma(\omega)\mathbf{E}$. P-polarization excites only densities $J_x$ and $J_z$ or E-wave, s-polarization leads to $J_y$ or H-wave. The tensor conductivity connects all three components of the current to the field. In a normal fall on the plate with $\alpha=0$ the wave with polarization of the electric vector is normal to the layers passes with much lower losses than the wave with orthogonal polarization. In the asymmetric case, this works for *p*-polarized and for *s*-polarized waves incident at a certain angle.

Below we obtain Fresnel formulas when *p*-polarized and *s*-polarized wave falling from below at the angle $\varphi$ on a structure with an arbitrary angle $\alpha$. We use of the impedance approach, considering the movement along the *x*-axis. In the bottom we have the wave

$$E_q = \exp(-ik_{0x}x) + R_q \exp(ik_{0x}x), \quad Z_0 H_q = [\exp(-ik_{0x}x) - R_q \exp(ik_{0x}x)]/\rho_{0q}.$$

From above we have

$$E_q = T_q \exp(-ik_{0x}(x-d)), \quad Z_0 H_q = T_q \exp(-ik_{0x}(x-d))/\rho_{0q}.$$

In structure we write

$$E_q = A_q^+ \exp(-ik_x^+ x) + A_q^- \exp(-ik_x^- x), \quad Z_0 H_q = A_q^+ \exp(-ik_x^+ x)/\rho_q - A_q^- \exp(-ik_x^- x)/\rho_q.$$

Here for $q = p$ we have $E_q = -E_z$, $H_q = H_y$, $\rho_{0p} = \sqrt{1 - k_z^2/k_0^2}$, and $\rho_q = \sqrt{(\tilde{\varepsilon}_{xx} - k_z^2/k_0^2)/\Delta}$. When $q = s$ we have $E_q = E_y$, $H_q = H_z$, $\rho_{0s} = 1/\sqrt{1 - k_z^2/k_0^2}$, $\rho_s = k_0/\sqrt{k_0^2 \tilde{\varepsilon}_{xx} - k_z^2}$. By matching the field components, one can obtain the solution:

$$T_q = \frac{4\rho_q \rho_{0q}}{\exp(ik_x^+ d)(\rho_q + \rho_{0q})^2 + \exp(ik_x^- d)(\rho_q^2 - \rho_{0q}^2)}, \tag{28}$$

$$A_q^+ = \exp(ik_x^+ d) T_q (1 + \rho_q/\rho_{0q})/2,$$

$$A_q^- = T_q \exp(ik_x^- d)(1 - \rho_q/\rho_{0q})/2.$$

Using this one can find the impedance

$$Z_q = (1 + R_q)/(1 - R_q) = \frac{A_q^+ + A_q^-}{(\rho_{0q}/\rho_q)(A_q^+ - A_q^-)},$$

and the reflection coefficient in the form $R_q = (Z_q - 1)/(Z_q + 1)$ or as $R_q = A_q^+ + A_q^- - 1$. It should be borne in mind that in these relations the magnitude $k_z < k_0$ and the real, and the angle of incidence is defined as $\phi = \arctan(k_z/\sqrt{k_0^2 - k_z^2})$. In the case of symmetry, the equation is simplified and takes the form

$$T_q = \frac{4\rho_q \rho_{0q} \exp(-ik_x d)}{(\rho_{0q} + \rho_q)^2 + \exp(-2ik_x d)(\rho_q^2 - \rho_{0q}^2)}. \tag{29}$$

### 6. Numerical results and discussion

Figure 2 shows the calculation results $R = R_e$ and $T = T_e$ depending on the angle of incidence for wave length $\lambda = 500$ nm. Note that in the theory of diffraction on lattices, such problems for infinitely thin perfectly conducting strips are reduced to integral equations and have been solved. It is possible to take into account the final impedance of metal tapes. However, accounting for the dielectric layer in this approach is complex and requires the introduction of combined volume-surface integral equations. The results of calculation of dispersion and losses for PP at $\alpha = 0$ and $\alpha = \pi/2$ are shown in Fig. 3 and 4. Parts of the curves in the left region separated by line $n' = 1$ correspond to fast gliding waves, and to the right region describe the slow ones. The waves in the lower plane-layered structure ($\alpha = \pi/2$) are slower, and in the region above the frequency of plasmon resonance are inverse. This distinguishes the plane-layered AM from the metal one, for which there are no backward waves. For HMM with $\alpha = 0$ there are no backward waves, and the maximum deceleration corresponds to higher frequencies for which $\varepsilon' > 0$. For convenience the curves are constructed so that $k'_z > 0$, therefore, the backward waves correspond to a kind of negative loss. In a vacuum energy is always transferred along the motion of the phase. The presence of backward waves is an integral effect associated with the fact that in metal structures the component of the Poynting vector can change the sign if $\varepsilon' < 0$.

In the region of plasmon resonance $k'_z \approx k''_z \sim 1/\sqrt{\varepsilon''}$, i.e. to obtain large decelerations, dissipation should be reduced. The estimation for the first structure gives in the resonance region $k_z = k_0(1-i)\sqrt{(1 + \varepsilon_d^{-2} - 2/\varepsilon_d)/\varepsilon''}/2$. For the second structure, there are two resonances $k_z = k_0(1-i)\varepsilon_d/\sqrt{(1 + \varepsilon_d^2)\varepsilon''}$ and $k_z = k_0(1+i)\sqrt{(1+\varepsilon_d^2 - \varepsilon_d)/[(1+\varepsilon_d^2)2\varepsilon'']}$, respectively, a low-frequency at $\varepsilon_m \approx -\varepsilon_d$ and a high-frequency at $\varepsilon_m \approx 1/\varepsilon_d$. It is immediately seen that the latter corresponds to the backward PP. All investigated PP are gliding, since the leakage from the dissipative half-space is impossible.

In general case we should iteratively solve the equation (13), presenting it, for example, in the form of

$$k_z^\pm = \pm k_0 \left[ \frac{\tilde{\varepsilon}_{xx} f(k_0, k_z)/\Delta - 1}{f(k_0, k_z)/\Delta - 1} \right]^{1/2}. \tag{30}$$

Here the even function of $k_z$ is denoted:

$$f(k_0, k_z) = \left[ \frac{(\rho - \rho_0)\exp(-ik_x^+ d) - (\rho + \rho_0)\exp(-ik_x^- d)}{(\rho - \rho_0)\exp(-ik_x^+ d) + (\rho + \rho_0)\exp(-ik_x^- d)} \right]^2.$$

Indeed, it can be represented as

$$f(k_0, k_z) = \left[ \frac{(\rho - \rho_0)\exp(-i\psi) - (\rho + \rho_0)\exp(i\psi)}{(\rho - \rho_0)\exp(-i\psi) + (\rho + \rho_0)\exp(i\psi)} \right]^2,$$

where $\psi = \pm d\sqrt{(k_0^2 \tilde{\varepsilon}_{xx} - k_z^2)\Delta}/\tilde{\varepsilon}_{xx}$. Two branches in (30) $k_z^+$ and $k_z^- = -k_z^+$ are related with $k_x^\pm(k_z)$ from (6), so that $k_x^+(k_z^+) = -k_x^-(k_z^-)$. Thus, the DE (30) defines two waves with opposite directions of phase velocities for mutual structure. Two values of $\psi$ give two dispersive branch of waves for any direction, one of which more slow then another. So, the total number of waves of both directions is four. Fig. 5 shows the results of the iterative solution of DE (30). Maximum deceleration increases with $\alpha$ increasing. For symmetrical cases $\alpha = 0$ and $\alpha = \pm\pi/2$ also there are two branches: symmetric and antisymmetric (with electric and magnetic walls in the center) which described by the DEs $\rho_0 = \rho[i\tan(k_x d/2)]^{\pm 1}$. We must propose, that $t_p << d$ to use the effective media approximation for the case $\alpha = \pm\pi/2$. This limits the thickness of the structure from below and, accordingly, the maximum deceleration of slower plasmon (with a magnetic wall) compared to the plasmon along a thin metal film [27,28].

## 7. Conclusions

In this paper, using the simplest homogenization, the exact solutions are obtained for PP along a layer in the general case of asymmetric HMM in the form of a plane-layered periodic metal-dielectric structure. An additional degree of freedom can be introduced into the equations by varying the fill factor, or by using multilayer structures in the period. Taking into account SD, i.e. dependence $\tilde{\varepsilon}(k_0, k_x, k_z)$, leads to complex nonlinear DE and Fresnel equations, which can only be analyzed numerically. Both SD and dissipation distort the hyperbolic law of dispersion and limit the modulus of the wave vector components, i.e. close the isofrequency surface [22]. The conditions for the existence of slow and fast, gliding and leakage, as well as forward and backward PPs are found. The asymmetric layer of HMM is interesting in that it supports PP with different conditions of gliding/leakage on both sides. The backward PPs were found along the plane-layered structure of the half-space, but they are absent along the metallic half-space. To reduce the frequency $\omega_p$ and plasmonic resonance frequency one can use the semiconducting layers instead metallic, and to reduce a losses and increase the wave deceleration the low temperature of graphene optical pumping HMMs are needed [38,43]. Cryogenic temperatures or active elements (spasers) can be used to reduce losses.

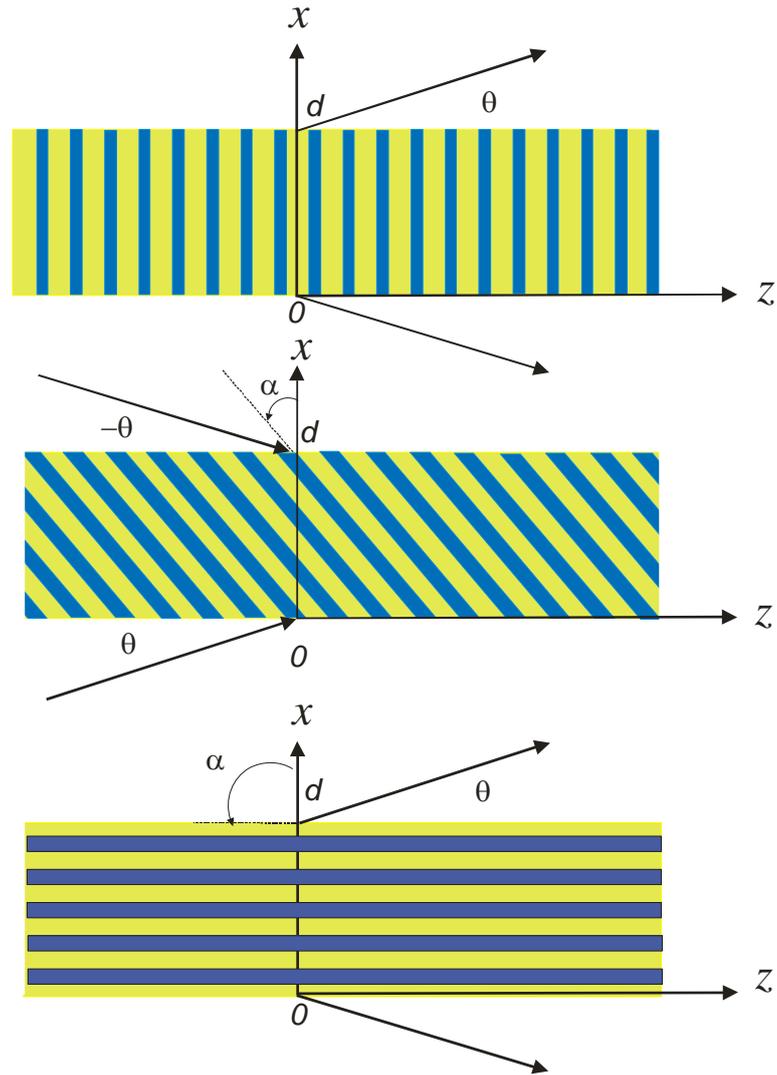

Fig. 1. The layer of HMM with thickness *d* consisting of conductive metal sheets periodically embedded in the dielectric (from above), and the same layers obtained by cuts the HMM to the crystallographic axis at angle α=π/4 (in center) and angle α=π/2 (below)

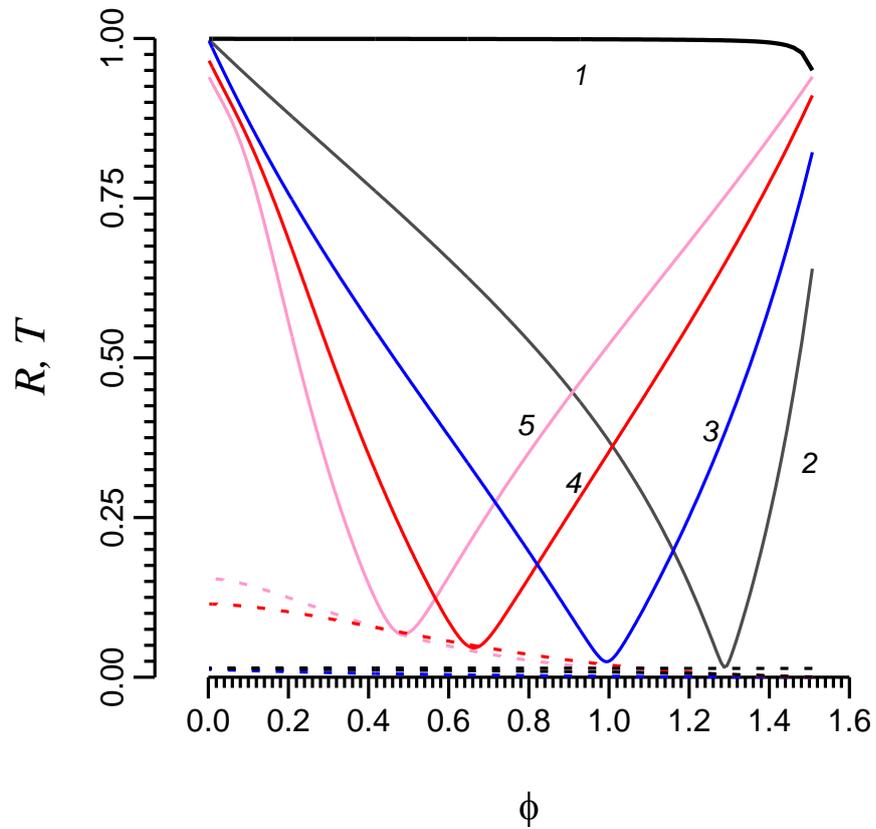

Fig. 2. The dependence of the reflectance $R$ (solid curves) and transmission $T$ (dashed curves) of the angle of incidence $\phi$ for the structure of HMM with $d=420$ nm $t_m=t_d=20$ nm, $\varepsilon_d =3$ at different values of angle α: α=0 (curve 1), α=π/12 (2), α=π/8 (3), α=π/4 (4), α=π/3 (5)

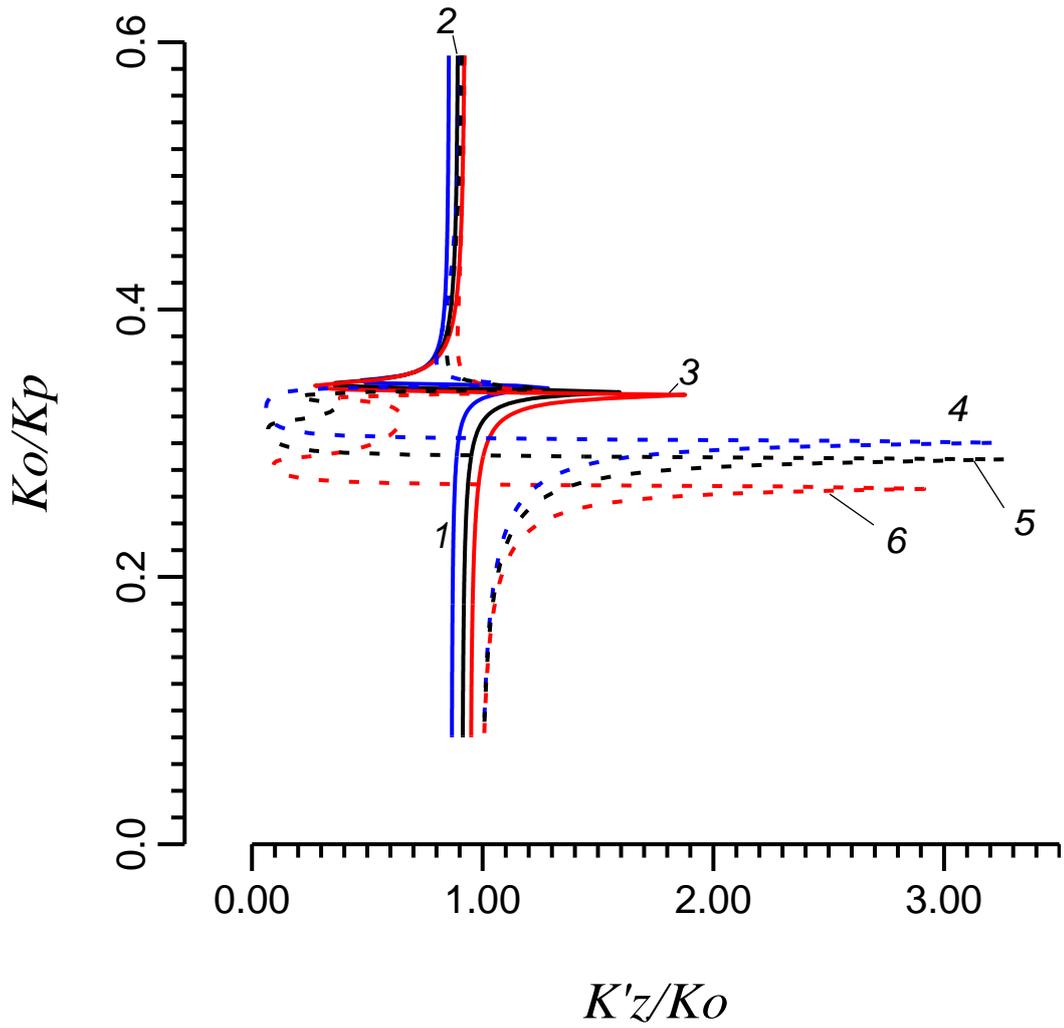

Fig. 3. E-PP dispersion along the half-space of the HMM with a structure corresponding to Fig. 2 for α=0 (curves 1–3) and α=π/2 (curves 4–6): the dependence of the normalized wavenumber on deceleration $n' = k'_z / k_0$ at different DP $\varepsilon_d$: 2.0 (curves 1–4), 3.0 (curves 2–5) and 5.0 (curves 3–6)

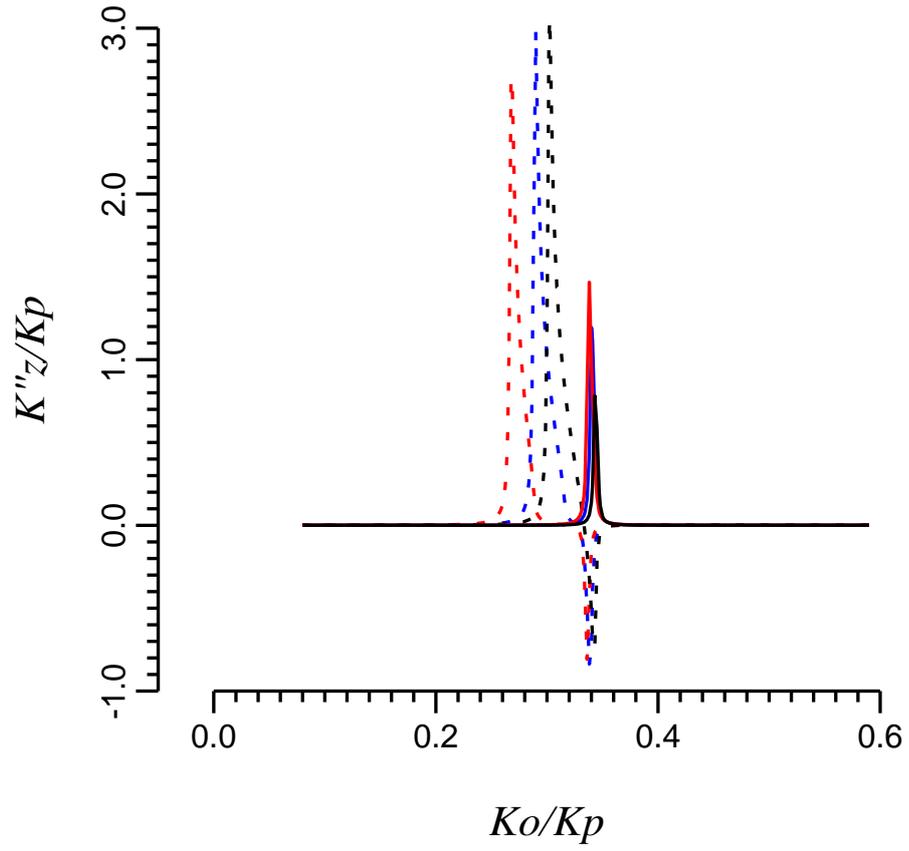

Fig. 4. Normalized losses $k_z''/k_p$ of E-PP along the half-space of HMM, corresponding to the dispersion of Fig. 2 for α=0 (curves 1–3) and α=π/2 (curves 4–6) and depending on the normalized wave number $k_0/k_p$ at different DP $\varepsilon_d$: 2.0 (curves 1–4), 3.0 (curves 2-5) and 5.0 (curves 3–6)

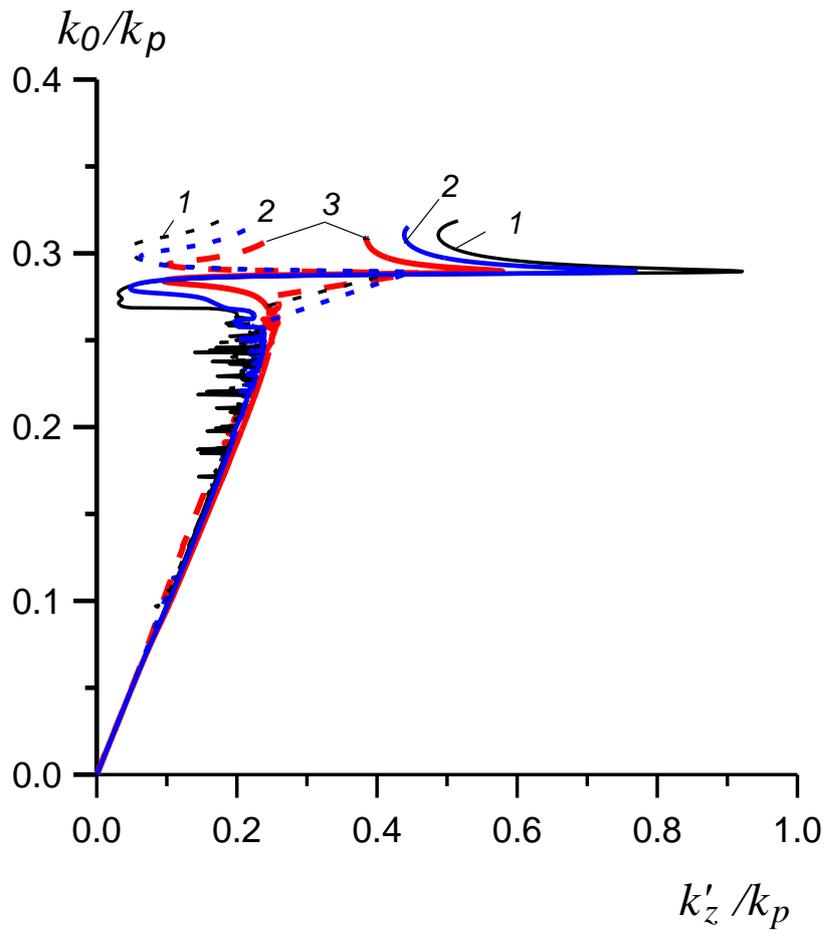

Fig. 5. Dispersion in a layer of asymmetric HMM of 50 nm thickness depending on $\alpha$: $\pi/3$ (curves 1), $\pi/4$ (2) and $\pi/6$ (3). Solid curves correspond to slower PP and the "−" sign in the expression for $\psi$, and dashed curves correspond to the "+" sign